# AN APPLICATION OF NEURAL NETWORKS TO CHANNEL ESTIMATION OF THE ISDB-T$_B$ FBMC SYSTEM

Jefferson Jesus Hengles Almeida, P. B. Lopes, Cristiano Akamine, and Nizam Omar

Postgraduate Program in Electrical Engineering and Computing
Mackenzie Presbyterian University, Sao Paulo, Brazil
jefferson.a.br@ieee.org
paulo.lopes@mackenzie.br
akamine@ieee.org
nizam.omar@mackenzie.br

## ABSTRACT

*Due to the evolution of technology and the diffusion of digital television, many researchers are studying more efficient transmission and reception methods. This fact occurs because of the demand of transmitting videos with better quality using new standards such 8K SUPER Hi-VISION. In this scenario, modulation techniques such as Filter Bank Multi Carrier, associated with advanced coding and synchronization methods, are being applied, aiming to achieve the desired data rate to support ultra-high definition videos. Simultaneously, it is also important to investigate ways of channel estimation that enable a better reception of the transmitted signal. This task is not always trivial, depending on the characteristics of the channel. Thus, the use of artificial intelligence can contribute to estimate the channel frequency response, from the transmitted pilots. A classical algorithm called Back-propagation Training can be applied to find the channel equalizer coefficients, making possible the correct reception of TV signals. Therefore, this work presents a method of channel estimation that uses neural network techniques to obtain the channel response in the Brazilian Digital System Television, called ISDB-T$_B$, using Filter Bank Multi Carrier.*

## KEYWORDS

*Channel estimation, Artificial intelligence, ISDB-T$_B$, FBMC, Neural Networks.*

## 1. INTRODUCTION

Digital TV standards in current use allow for the transmission of standard or high definition video content. However, consumers are demanding more resolution for more realistic experiences while watching TV. Therefore, researchers all over the world are working in the development of the concepts that will enable the broadcast of ultra-high definition content. These concepts include novel modulation schemes, powerful channel estimation, intelligent receptors, antenna arrays, etc.

Filter Bank Multi Carrier (FBMC) is a modulation technique that has been applied as an alternative to the Orthogonal Frequency Division Multiplexing (OFDM) [1]. This trend is due to the fact that FBMC does not use the Cyclic Prefix (CP), increasing significantly the system data rate [2]. Thus, when FBMC is applied to digital television system such as Integrated Services Digital Broadcasting Terrestrial B (ISDB-T$_B$) associated with channel coding and synchronization techniques such as Low Density Parity Check (LDPC) and Bootstrap, 8K transmission may be made viable. Nevertheless, the transmission of more information per second using the same frequency bandwidth leads to an increase in bit error rate if improved channel estimation algorithms, powerful error correcting codes and novel equalizers are not used. However, the channel estimation becomes a little more complex, due to the characteristics of the used filters and degrading components present on the channel [3]. In this article, an intelligent channel estimation algorithm using Neural networks is reported.

The channel estimation is a crucial stage for the perfect reception of digital TV signals because of the interferences that are generated by several sources on the channel. In the specific case of the terrestrial broadcast, these are Additive White Gaussian Noise (AWGN), multipath, and others [4]. Therefore, it is necessary to use different techniques and processes that make possible the removal and minimization of these effects, allowing the reception of the transmitted signal in an appropriated way.

Several techniques can be used to estimate the frequency response of the channel. Among them, we can highlight those that use pilots associated to interpolation methods that adequately minimize the AWGN and multipath effect [5]. However, this process is not always trivial, and can be improved with the use of Artificial Intelligence (AI).

The AI can be understood as a set of algorithms to solve complex problems, being able to resolve, to make decisions, and to develop a method of learning, according to the situation to which it is applied [6]. In this context, Neural Networks (NN) are used to solve problems through the simulation the connection between brain neurons, using specific activation and training algorithms [7].

The herein proposed channel estimation method uses a NN, trained through the Back-propagation algorithm to calculate the channel response and to permit the correct equalization in a scenario with AWGN and multi paths. This technique is applied to an FBMC version of the ISDB-$T_B$ digital TV standard. The overall system is simulated on GNURadio environment. The presented results show the feasibility of this new system even when transmission is performed in a channel with severe multipath interference.

## 2. ISDB-$T_B$

The Brazilian Digital Television System (SBTVD), ISDB-$T_B$, is the terrestrial digital TV standard adopted by 18 countries in the world. It employs a bandwidth of 6, 7, or 8 MHz and can transmit One-Segment (1SEG), Standard Definition Television (SDTV) and High Definition Television (HDTV), according to the combination chosen for the 13 Segments available. In Figure 1 it is possible to see the bandwidth segmentation used in 6 MHz. The 3 modes of operation have different parameters that are detailed in Table I [8].

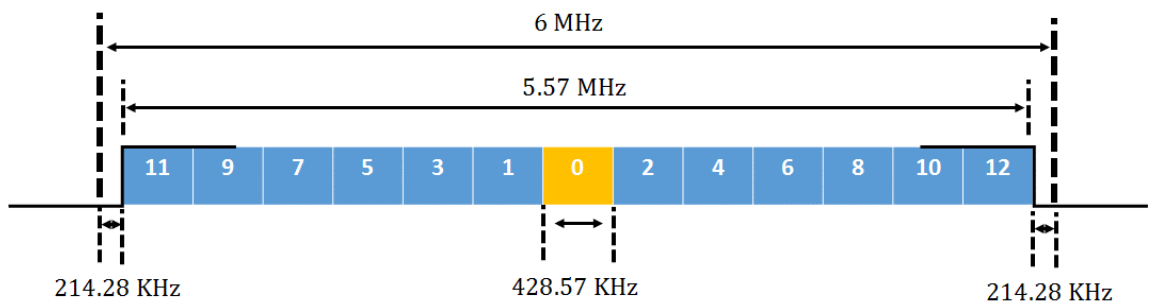

Figure 1. Segmentation of the ISDB-$T_B$ channel.

Table 1. ISDB-T$_B$ transmission parameters.

| Mode | Parameters | | | | |
|---|---|---|---|---|---|
| | Carriers | Useful Carriers | Pilots | Symbol period | IFFT |
| 1 | 1405 | 1248 | 157 | 0,252 ms | 2048 |
| 2 | 2809 | 2496 | 313 | 0,504 ms | 4096 |
| 3 | 5617 | 4992 | 625 | 1,008 ms | 8192 |

The modulations *Quadrature Phase Shift Key* (QPSK), Diferencial QPSK, 16-QAM e 64-QAM can be used. The constellations of three are presented in the Figure 2.

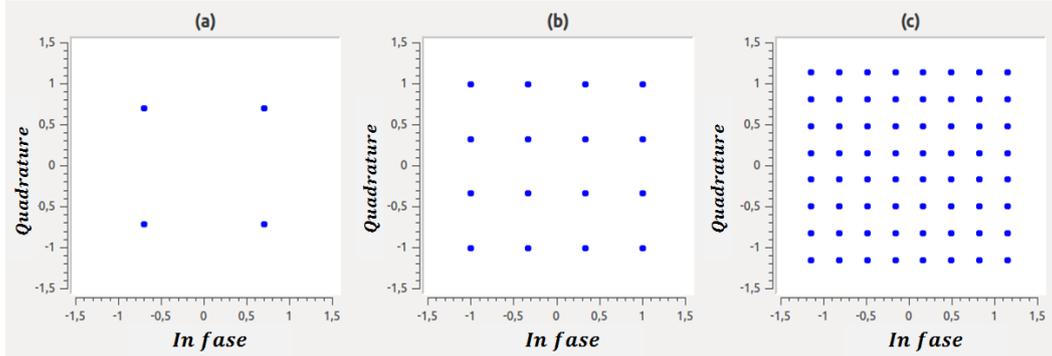

Figure 2. Constellations of (a)QPSK, (b)16-QAM, and (c)64-QAM.

The CP values can be 1/4, 1/8, 1/16, or 1/32 of the useful time symbol.

## 3. FBMC

The Filter Bank Multi Carrier (FBMC) modulation technique consists of dividing the available bandwidth into small equally spaced small segments [9] (Figure 3).

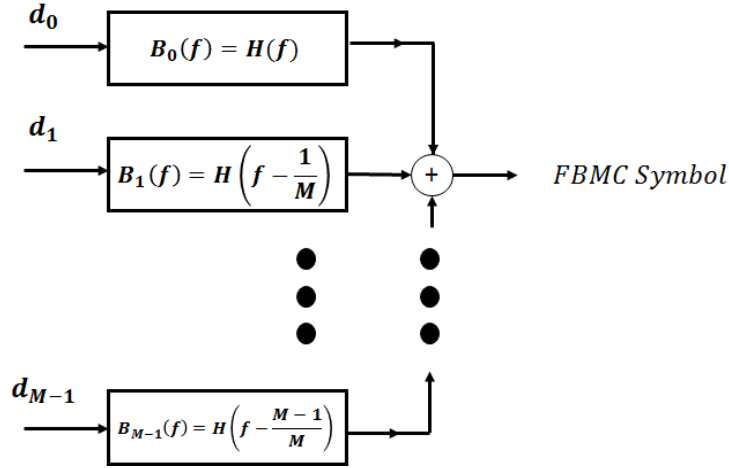

Figure 3. FBMC composed symbol.

For this purpose, filter banks complying with (1) are used.

$$B_k(f) = H\left(f - \frac{k}{M}\right) = \sum_{i=0}^{L-1} h_i e^{-j2\pi i\left(f - \frac{k}{M}\right)} \qquad (1)$$

where, $f$ is the frequency, $M$ is the number of subcarriers, $L$ is the number of filter coefficients, and $k = 0 \dots M-1$.

Applying the Z transform in (1) it is found (2).

$$B_k(z) = \sum_{i=0}^{L-1} h_i e^{2j\pi i \frac{k}{M}} z^{-i} \quad (2)$$

After the polyphase decomposition, (3) can be observed.

$$B_k(z) = \sum_{p=0}^{M-1} e^{2j\pi \frac{k}{M} p} z^{-p} H_p(z^M) \quad (3)$$

Finally, making $W_M = e^{-\frac{2j\pi}{M}}$, (4) can be derived.

$$B_k(z) = \sum_{p=0}^{M-1} W_M^{-kp} z^{-p} H_p(z^M) \quad (4)$$

By expressing this equation in matrix notation, (5) is obtained.

$$\begin{bmatrix} B_0(z) \\ \vdots \\ B_{M-1}(z) \end{bmatrix} = \begin{bmatrix} 1 & \cdots & 1 \\ \vdots & \ddots & \vdots \\ 1 & \cdots & W_M^{-(M+1)^2} \end{bmatrix} \begin{bmatrix} H_0(z^M) \\ \vdots \\ z^{-(M-1)} H_{M-1}(z^M) \end{bmatrix} \quad (5)$$

The implementation of this equation is depicted in the block diagram in the Figure 4.

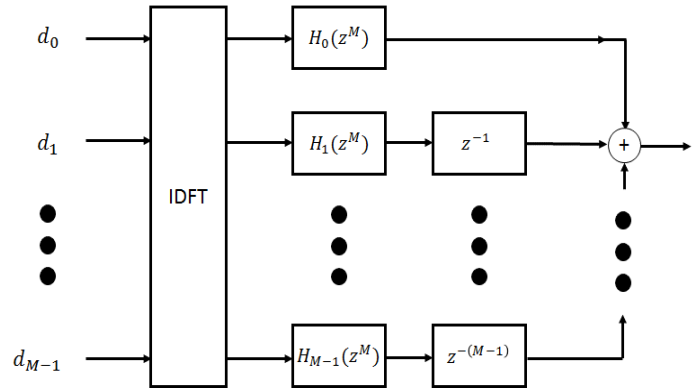

Figure 4. FBMC system model.

### 3.1. Designed Filters

The filters are designed according to the zero inter-symbol interference Nyquist criterion to avoid phase and amplitude distortions [10]. Thus, using $K = 4$, the criteria showed in (6) and (7) are employed.

$$h_3 + h_2 + h_1 + h_0 + h_1 + h_2 + h_3 = 0 \quad (6)$$

$$\begin{aligned} h_0 &= 1 \\ h_1^2 + h_3^2 &= 1 \\ 2h_2^2 &= 1 \end{aligned} \quad (7)$$

Then, the frequency coefficient values are calculated and presented in (8)

$$h_0 = 1$$
$$h_1 = -0.9719598$$
$$h_2 = 0.7071068$$
$$h_3 = -0.2351470$$
(8)

The frequency response of one subcarrier can be seen in the Figure 5.

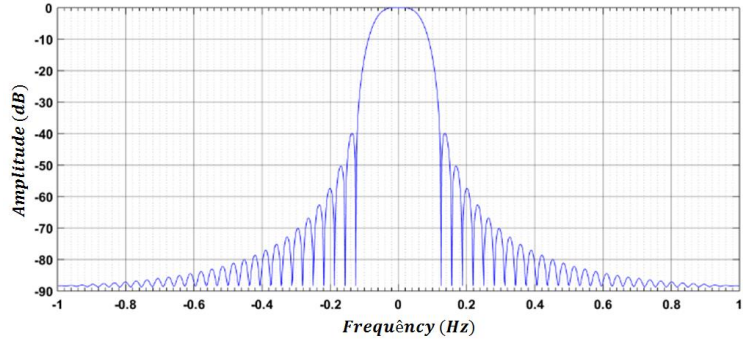

Figure 5. Frequency Response.

From the Inverse Fourier Transform, it is possible to analyse the time impulse response, such as in Figure 6.

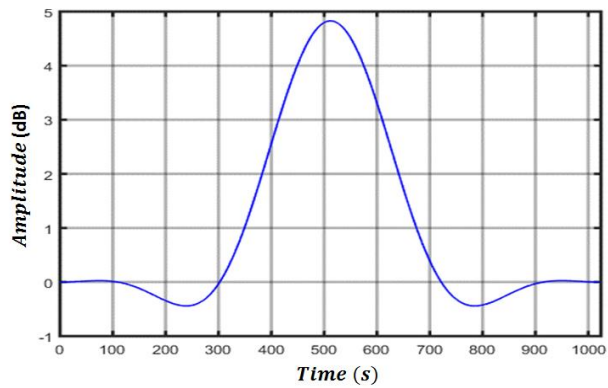

Figure 6. Time Impulse Response.

### 3.2. Modulation OQAM

FBMC employs Offset Quadrature Modulation (OQAM), so that the orthogonality is obtained between symbols and not between subcarriers [11]. At the transmitter, the pre processing block (Figure 7) separates the complex symbol in two purely real parts in a staggered way.

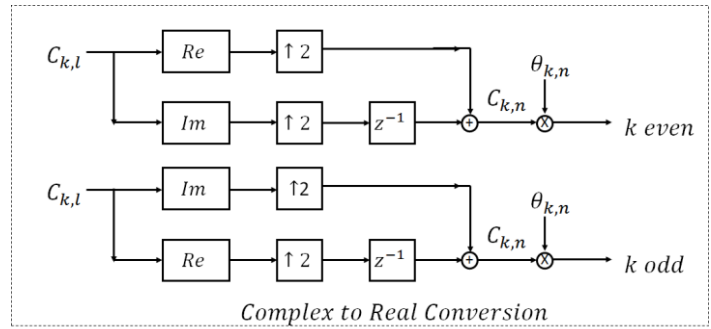

Figure 7. Pre processing OQAM.

At the receiver the post processing block (Figure 8) accomplish the junction of the real parts in the original complex symbol. Thus, it is not necessary to use the CP, making it possible to increase the data rate of the system.

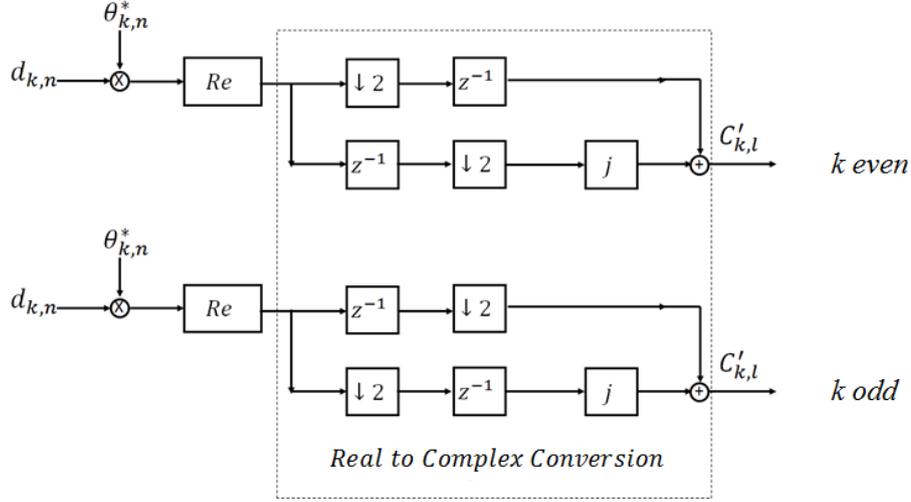

Figure 8. Post processing OQAM.

### 3.3. Beta Multiplier

To make the FBMC system to be causal, it is necessary to multiply all the subcarriers by a complex factor called Beta Multiplier [12]. The, using the number of subcarriers $M$ and the filter length $L_p = KM - 1$, (9) is applied.

$$\beta_k = (-1)^k e^{-j\pi k \frac{L_p - 1}{M}} \quad (9)$$

## 4. PILOT BASED CHANNEL ESTIMATION

To estimate the frequency response of the channel, ISDB-$T_B$ performs the constant transmission of pilots. The position of the pilots depends on a Pseudorandom Binary Sequence (PRBS) sequence that has a generator polynomial equal to $x^{11} + x^9 + 1$ [8].

After the transfer function ($Hp$) is found using (10), where $Y_p(m)$ and $X_p(m)$ are the pilot received and transmitted signals through the m$^{th}$ subcarrier, an interpolation method that can be linear, cubic, among others, is used to estimate the responses at the frequency corresponding to the k$^{th}$ subcarrier, located between the m$^{th}$ and the (m+1)$^{th}$ pilot subcarriers.

$$H_p(m) = \frac{Y_P(m)}{X_p(m)} \quad (10)$$

In the case of linear estimation, we use (11).

$$H(k) = (1 - a) \cdot H_p(m) + a \cdot H_p(m + 1) \quad (11)$$

where $a$ is a constant determined by the relation between the distance of the position of the subcarrier where it is desired to estimate the response of the channel to the position of the nearest pilot.

In the case of cubic interpolation, (12) is used.

$$H(k) = A(a) \cdot H_p(m) + B(a) \cdot H_p(m + 1) + C(a) \cdot z(m) + D(a) \cdot z(m + 1) \quad (12)$$

where $A(a)$, $B(a)$, $C(a)$, and $D(a)$ are constants related to a and $z(m)$ is the second derivative obtained from the pilot information matrix [13].

## 5. NEURAL NETWORKS

Neural networks can be understood as algorithms that seek to simulate the functioning of the human brain, starting from the construction of small computational entities that act as a human neuron [14]. To do so, we use units called perceptrons (Figure 9) which have as input parameters an input $(x)$, a gain $(w)$, an activation function $(f)$ and an output $(y)$.

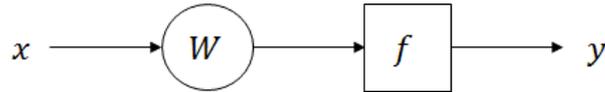

Figure 9. Perceptron.

The activation function can be linear or not depending on the desired application. The most common are the logarithm and the sigmoid. Perceptrons can be combined to form layers that are interconnected to generate larger and more complex networks. For the network to work correctly, it is necessary to perform the network training, using a set of known inputs and outputs, so that the gains are properly adjusted and the actual inputs generate the desired responses [15]. Among the training techniques, one of the most used is the Back-propagation Training Algorithm.

### 5.1. Back-propagation Training Algorithm

This technique uses a generalization of the Least Mean Square [16]. The activation function is defined as (13).

$$f(\alpha) = \frac{1}{1-e^{-\alpha}} \qquad (13)$$

Initially random weights $(w)$ are defined for the inputs $(x)$. Then from the desired response $(d)$ the error is calculated by (14).

$$erro = 0.5 \cdot (d-y)^2 \qquad (14)$$

where $y$ is the output of the activation function.

Then the weights are updated, using (15).

$$w(t+1) = w(t) - \delta \cdot erro \cdot x \qquad (15)$$

where $t$ is the previous iteration of the algorithm has been set and $\delta$ is a chosen gain. Finally, the algorithm is repeated until the desired response is obtained at the output of the system and the weights are properly adjusted.

## 6. PREVIOUS DEVELOPED MODEL OF ISDB-$T_B$ USING FBMC

In [2] and [17], a modified ISDB-$T_B$ system using FBMC was developed, using GNU Radio Companion (GRC) as simulation environment. The channel estimation algorithm used did not employ any Artificial Intelligence feature. For this reason, the present work expands those articles by using a different approach based on Neural Networks.

### 6.1. GRC

The GRC is a computational tool that allows the development of processing blocks for simulating communications systems [18]. It is an open source and free software that makes possible the interface between the created model and software radio peripherals [19]. It uses a Graphical User Interface (GUI) that facilitates the software handling [20] [ 21].

The block codes are created using the C/C++ or Phyton languages and the interconnection among these blocks is described only in Phyton [22].

The processed data sources on GRC are of complex (8 bytes), float (4 bytes), int (2 bytes), or byte (1 byte) types. The used terminology of GRC is presented on Table II [23].

Table 2. GRC terminology.

| Name | Definition |
| --- | --- |
| Block | Processing Unit with ins or outs |
| Port | Block input or output |
| Source | Data generator |
| Sink | Data consumer |
| Connection | Data flow from an output to an input |
| Flow Graph | Set of blocks and connections |
| Item | Data unit |
| Stream | Continuous flow of items |
| IO Signature | In and out description |

Then using GRC and programming language the system can be simulated.

### 6.2. Flow Graph of ISDB-T$_B$ FBMC

The transmitter implemented on GRC is presented in the diagram shown on Figure 10.

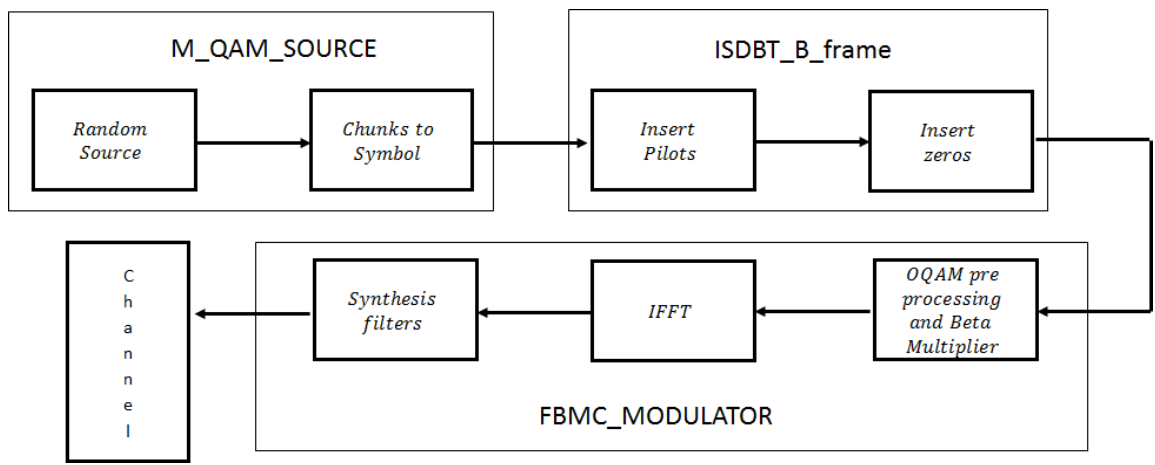

Figure 10. ISDB-T$_B$ FBMC transmitter.

As it can be seen at the transmission side, an information source generates data that is modulated, formatted according to the standard, processed by OQAM pre-processing, multiplied by Beta, modulated through the IFFT and Synthesis filters and transmitted through the channel.

The receiver diagram is shown in the Figure 11.

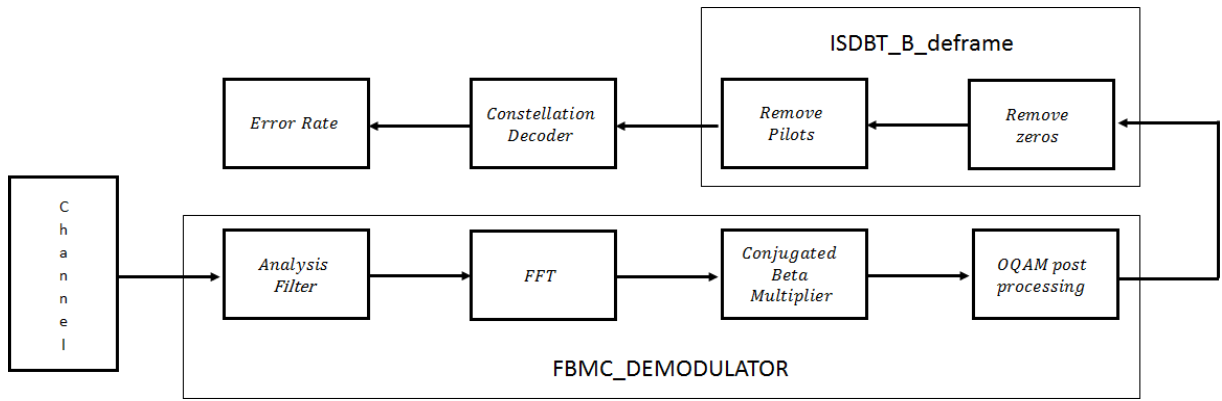

Figure 11. ISDB-$T_B$ FBMC receiver.

At the reception, the data goes through the analysis filters and FFT, multiplied by conjugated Beta, processed in OQAM post processing. After the zeros and pilots are removed, the data is decoded to calculate the Bit Error Rate (BER).

The GRC environment flow graph is depicted in Figure 12.

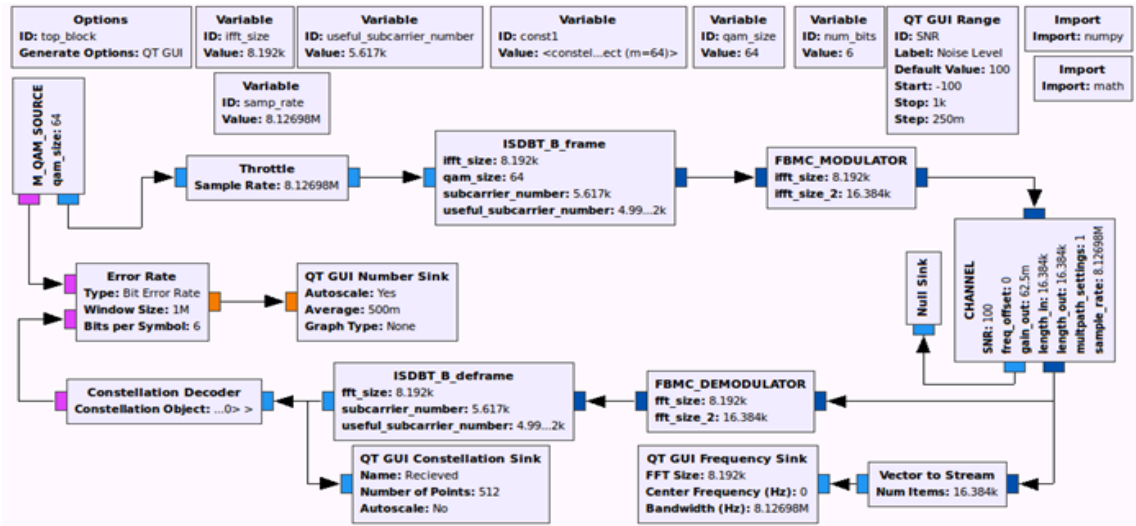

Figure 12. Flow Graph of ISDB-$T_B$ FBMC.

## 7. PROPOSED IA ESTIMATION METHOD

To accomplish the AI estimation, a simple NN with one perceptron for each real and imaginary part of received symbol is used. When the system initiates, four FBMC symbols are sent as training sequence and the weights of NN are trained using the Back-propagation method. Then regular operation starts and the received data symbols are equalized by the trained system.

The Flow Graph used is shown in Figure 13. Inside the "ISDBT_B_deframe" hierarchical block, three different channel estimators were implemented: the linear interpolation and the cubic interpolation, both at the time and frequency, and a neural network estimator trained with the Back-propagation technique (Figure 13).

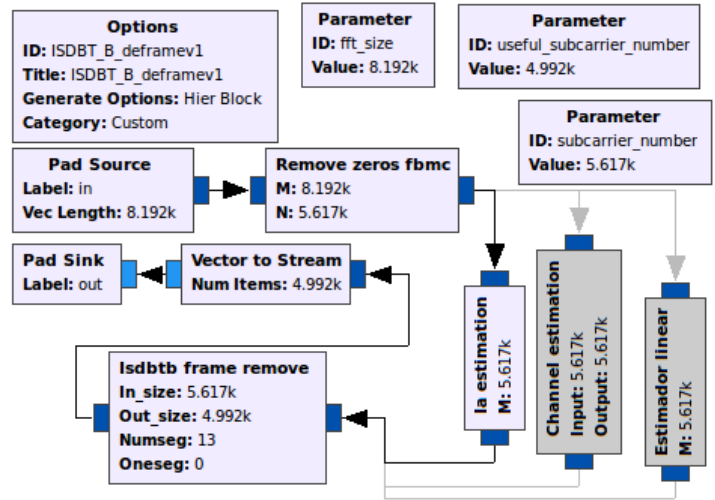

Figure 13. Content of the hierarchical block "ISDBT_B_deframe".

Thus, it was connected each channel estimator ate the system and the results could be collected.

## 5. RESULTS

The analysis was made using the ISDB-$T_B$ FBMC in mode 3 as in Table I. The Bit Error Rate (BER) curves were observed on two scenarios. The first is characterized by AWGN and modulation level equal to 64 (64-QAM) or 6 bits per symbol (Figure 14). The second uses the Brazil A digital TV channel model [24], which applies 6 paths with 0, 0.15, 2.2, 3.05, 5.86, and 5.93 microseconds of delay and 0, 13.8, 16.2, 14.9, 13.6, and 16.4dB of attenuation respectively, using the modulation level 4 (4-QAM) (Figure 15). In this last case, it was necessary to reduce the QAM modulation due to the long time required to perform the real-time simulation.

It can be observed that the use of neural networks brought to the system an increase of robustness at $10^{-5}$ BER level around to $2.1 dB$ in the case only of AWGN and around to $2 dB$ in the case which there is AWGN and multipath, when the comparisons are made considering cubic estimation.

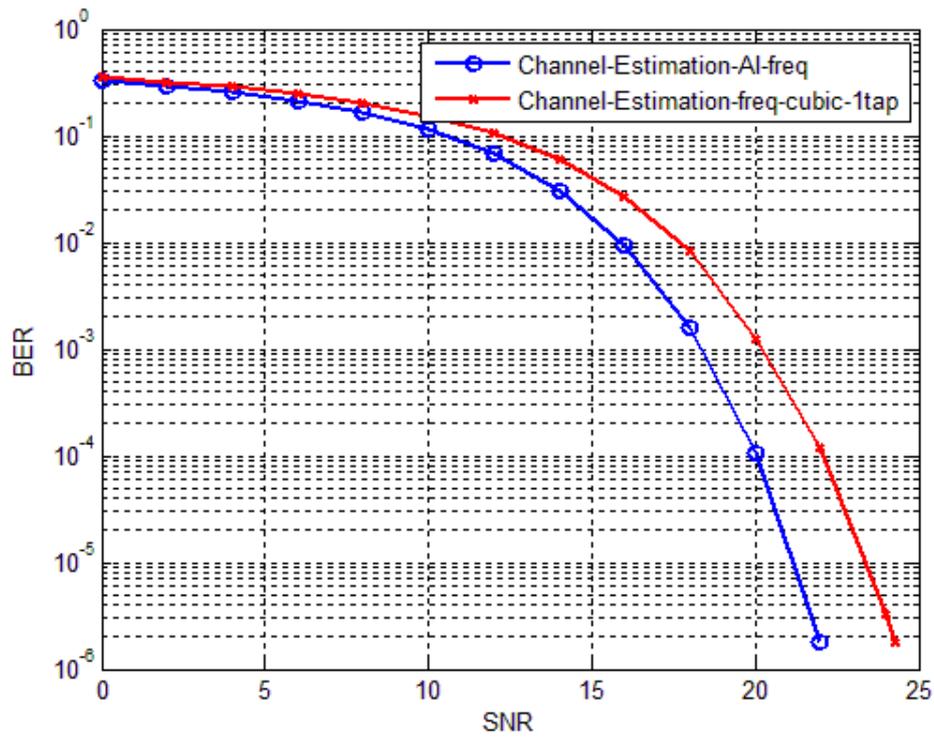

Figure 14. BER curves of ISDB-$T_B$ FBMC using 64-QAM with AWGN inserted.

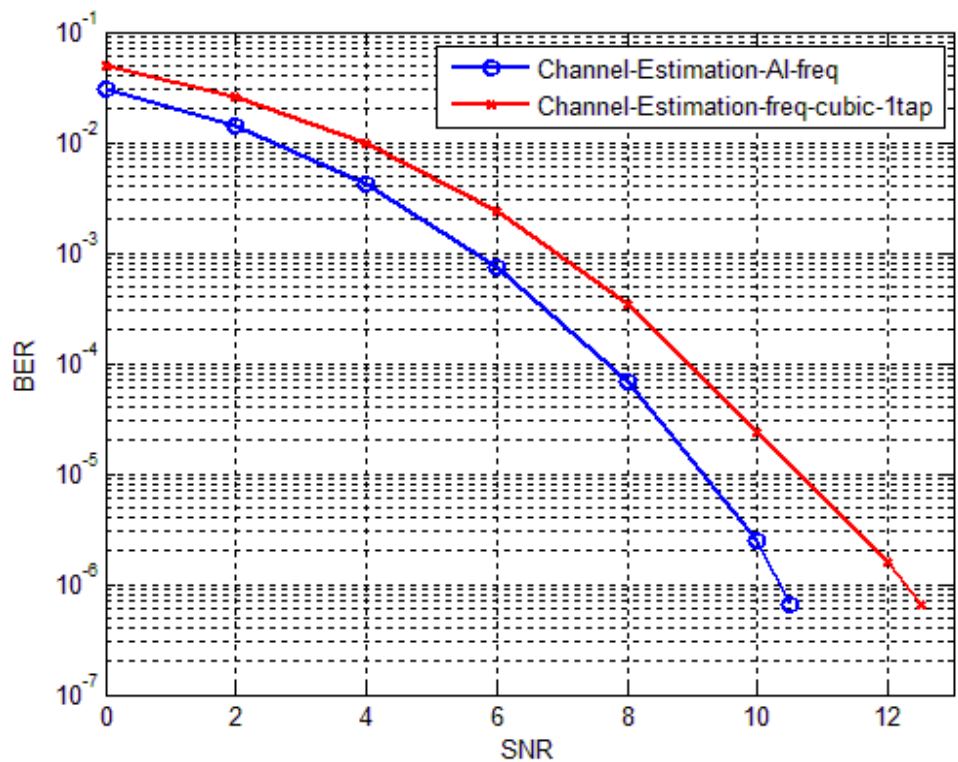

Figure 15. BER curves of ISDB-$T_B$ FBMC using 4-QAM with AWGN and multipath inserted.

# 6. Conclusion

Current digital TV standards were established to enable the broadcast of standard or high definition video. Nevertheless, nowadays consumers are demanding even higher definition content. For this reason, researchers are working on new standards that will enable a higher information transmission rate in terms of bits/s/Hz. This is the case of FBMC which was also proposed for forthcoming 5G cellular standard. But this increase in bit rate on the same frequency bandwidth implies in an increase in bit error rate if novel channel estimation and equalization are not created.

The use of artificial intelligence applied to channel estimators for FBMC opens a new field of research. The possibility of employing smart algorithms that can learn even in the presence of nonlinear interference is paramount to the success of more spectrally efficient modulation techniques.

In this paper, it was showed that the application of a simple neural network to the problem of channel estimation in the FBMC modified ISDB-Tb digital TV standard is feasible. The presented technique achieved an increase in 15% in robustness in a channel with several multipath interferences, when the cubic method is compared to AI estimation. It was also shown that the Back-propagation training algorithm allows the estimation of the channel frequency response and contributes to minimize the bit error rate.

In future, other kind of neural networks will be investigated, such as a recursive network, since it can improve the results and required computing effort.

## Acknowledgements

The authors would like to thank the MACKPESQUISA, Coordination for the Improvement of Higher Level Personnel (CAPES) and National Counsel of Technological and Scientific Development (CNPq) for the partial financial subsides for this research.

**Authors**

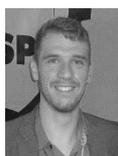

**Jefferson Jesus Hengles Almeida** was born in Cotia, on May 1992. Received his B.Sc. degree in Electrical Engineering from Mackenzie Presbyterian University, São Paulo, Brazil, in 2014. Received his M.Sc Degree in Electrical Engineering from Mackenzie Presbyterian University, São Paulo, Brazil, in 2016. He is currently studying his Ph.Ddegree in Electrical Engineering in Mackenzie Presbyterian University. His current research involves broadcasting areas, digital television transmission systemsstudies,and software defined radio.

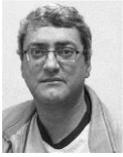 **Paulo Batista Lopes** was received the B.Sc. and M.Sc. in EE from the Universidade Federal do Rio de Janeiro, Brazil, in 1978 and 1981, respectively, and the Ph.D. in EE from Concordia University, Montreal, Canada, in 1985. From 1985 to 1988, he was with Elebra and CMA, two Brazilian companies, working on the design of several Communication equipments. From 1988 to 1999, he was with Texas Instruments as a DSP specialist. In 1999, he moved to Motorola-SPS (later to become Freescale Semiconductor) as a Sales and Application manager. Since 2009, he has been with Universidade Presbiteriana Mackenzie as a professor in the School of Engineering. His research interests are Circuit Theory, Digital Signal Processing, and Analog Circuit Design.

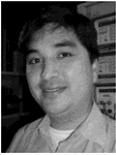 **Cristiano Akamine** received his B.Sc. degree in Electrical Engineering from Mackenzie Presbyterian University, São Paulo, Brazil, in 1999. He received his M.Sc. and Ph.D. degree in Electrical Engineering from the State University of Campinas (UNICAMP), São Paulo, Brazil, in 2004 and 2011 respectively. He is a professor of Embedded Systems, Software Defined Radio and Advanced Communication Systems at Mackenzie Presbyterian University. He has been a researcher in the Digital TV Research Laboratory at Mackenzie Presbyterian University since 1998, where he had the opportunity to work with many digital TV systems. His research interests are in system on chip for broadcast TV and Software Defined Radio.

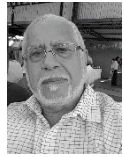 **Nizam Omar**, Mechanical Engineer ITA 1974. Master in Applied Mathematics, ITA 1979, Ph.D. in Computer Science - PUC RIO 1989. He is Professor at the Mackenzie Presbyterian University in Artificial Intelligence and its applications in Education, Engineering and Economics.